\documentclass{ifacconf}

\usepackage{graphicx}      
\usepackage{natbib}        

\usepackage{amsmath,amssymb}
\usepackage{mathtools}

\newtheorem{remark}{Remark}

\usepackage{algorithm}
\usepackage{algpseudocode}

\begin{document}
\begin{frontmatter}

\title{Closed-Loop Consistent, Causal Data-Driven Predictive Control\\via SSARX} 

\thanks[footnoteinfo]{This work was supported by VINNOVA Competence Center Dig-IT Lab Sustainable Industry, contract [2023-00556].}

\author[First]{Aihui Liu and Magnus Jansson} 

\address[First]{Department of Information Science and Engineering\\ 
KTH Royal Institute of Technology, 
   Stockholm, Sweden. \\
   (e-mail: aihui@kth.se; janssonm@kth.se).}

\begin{abstract}                
We propose a fundamental-lemma-free data-driven predictive control (DDPC) scheme for synthesizing model predictive control (MPC)-like policies directly from input–output data. Unlike the well-known DeePC approach and other DDPC methods that rely on Willems’ fundamental lemma, our method avoids stacked Hankel representations and the DeePC decision variable $g$. Instead, we develop a closed-loop consistent, causal DDPC scheme based on the multi-step predictor Subspace-ARX (SSARX). The method first (i) estimates predictor/observer Markov parameters via a high-order ARX model to decouple the noise, then (ii) learns a multi-step past-to-future map by regression, optionally with a reduced-rank constraint. The SSARX predictor is strictly causal, which allows it to be integrated naturally into an MPC formulation. Our experimental results show that SSARX performs competitively with other methods when applied to closed-loop data affected by measurement and process noise.
\end{abstract}


\end{frontmatter}






\section{Introduction}

Recent direct data-driven control methods, such as those in the Data-enabled Predictive Control (DeePC) framework in \cite{Coulson2019}, avoid explicit model identification by optimizing trajectories consistent with historical input–output data. These methods have seen rapid developments, including the maximum likelihood estimation (\cite{Yin2023}) and robust or chance-constrained formulations that explicitly address uncertainty (\cite{Coulson2022, Huang2023, Mattsson2023}). Some challenges remain active topics, such as process noise (\cite{Breschi2023}), causality (\cite{Sader2025}), and closed-loop issues (\cite{innovation, IVDDPC, Mejari2025}).

At the heart of DeePC lies Willems’ Fundamental Lemma (\cite{Willems2005}): for LTI systems, if an input is persistently exciting of sufficiently high order, then all future trajectories can be expressed by a linear combination of columns of a data Hankel matrix. This behavioral statement justifies replacing explicit predictors with Hankel data constraints. However, it also implies a heavy reliance on the data’s excitation and on regularization choices when noise is present (\cite{Dorfler2023,Breschi2023}).

In parallel, the system identification community developed subspace-based techniques, often aiming to identify explicit state-space models. A first step of such methods is fitting the data into a predictor-form equation, see e.g. \cite{finiteSIM}. This predictor equation in subspace identification can be embedded into MPC-like finite horizon optimizations, which connects subspace identification to data-driven predictive control. The early work on Subspace Predictive Control (SPC) in \cite{SPC} and its closed-loop version CLSPC in \cite{CLSPC} established connections between subspace identification and MPC. 

Recent results further make this bridge explicit in DeePC/ SPC or the direct/indirect control approaches. \cite{Dorfler2023} and \cite{Mattsson2024} show that regularization and relaxation connect DeePC and SPC predictors under certain conditions. Expressions for the implicit predictor in DeePC are derived in \cite{Kldtke2023}, which also indicates a close relation to the SPC predictor. The instrumental-variable reformulation of DeePC in \cite{Dinkla2026,Mejari2025} has made this bridge explicit by proving a direct equivalence between IV-DeePC and CLSPC, and by clarifying how instruments mitigate noise-induced bias.

This perspective motivates us to revisit the SPC-type multi-step predictors. Fundamental Lemma-based DeePC and SPC-style predictors both provide a similar finite-horizon input-output behavior: they both impose linear relations among stacked past/future data and do not require a parametric state-space model. DeePC uses Hankel-matrix coordinates, while SPC uses predictor/observer Markov-parameter coordinates estimated from data. 

With the decades of developments in subspace identification, these predictor-based methods are well understood statistically, have explicit bias/variance analysis, and have closed-loop variants with consistency guarantees; see. e.g. \cite{JANSSON1998,Jansson2000,finiteSIM} and the references therein.

Building on this, we adopt a classic subspace identification approach called Subspace-ARX (SSARX) in \cite{SSARX} for predictive control. SSARX, used as the identification step for constructing a multi-step predictor, does not rely on Willems’ fundamental lemma. SSARX still requires persistently exciting data, but it provides a natural framework for handling causality, process noise, and closed-loop data. It separates noise decoupling (ARX) from multi-step regression, mitigating tracking bias, which is unsolved in SPC (\cite{bias}). We also provide an optional low-rank regression, which can impose order constraints without explicitly identifying a state-space model.

The main contributions of this paper are:

(i) \textbf{A fundamental-lemma-free DDPC formulation.} 
We show that the input-output trajectory linear relationship can be built directly on a multi-step SSARX predictor, without using Willems’ fundamental lemma or an auxiliary trajectory variable $g$.

(ii) \textbf{A closed-loop consistent, causal predictor construction.} 
We present the long-existed SSARX algorithm for the non-parametric identification steps of the SSARX predictor, and discuss its properties.

(iii) \textbf{Empirical comparison and bias/variance analysis.} 
We demonstrate that SSARX-DDPC removes the SPC-type closed-loop bias in control and performs competitively with other methods. We further analyze the impacts of process and measurement noise levels in some DDPC methods.


Section 2 gives notations and reviews background on multi-step predictors in subspace identification. Section 3 presents the SSARX method, its low-rank variant, and discusses the closed-loop consistency result. A numerical experiment is carried out in Section 4, which includes the control cost comparison and a bias/variance analysis. Section 5 concludes the paper with some final remarks and limitations.

\section{Problem setting}
We consider the following discrete-time linear time-invariant (LTI) system in innovations form:
\begin{equation} \label{eq:inno_form}
\begin{cases}
x(t+1) = A x(t) + B u(t) + K e(t), \\
y(t) = C x(t) + D u(t) + e(t), 
\end{cases}
\end{equation}
where $x(t) \in \mathbb{R}^{n},u(t) \in \mathbb{R}^{n_u}$ and $y(t) \in \mathbb{R}^{n_y}$ denote the state, input and output signals, and $e(t)$ is the zero-mean white noise innovation.  

We will use the predictor form of (\ref{eq:inno_form}) by defining $\tilde{A}=A-KC$ and $\tilde{B}=B-KD$:
\begin{equation} \label{eq:pred_form}
\begin{cases}
x(t+1) = \tilde{A} x(t) + \tilde{B} u(t) + K y(t), \\
y(t) = C x(t) + Du(t)+ e(t) .
\end{cases} 
\end{equation}

We define the vectors of stacked outputs $\mathbf{y}_p (t) \in \mathbb{R}^{n_y L_p} $ and $\mathbf{y}_f (t) \in \mathbb{R}^{n_y L_f}$ of the past data length $L_p$ and the future data length $L_f$ as
\begin{align}
    \mathbf{y}_p(t) &= [y^\top(t-L_p), y^\top(t-L_p+1), \dots, y^\top(t-1)]^\top  \\
    \mathbf{y}_f(t) &= [y^\top(t), y^\top(t+1), \dots, y^\top(t+L_f-1)]^\top 
\end{align}
and the stacked inputs $\mathbf{u}_p (t) \in \mathbb{R}^{n_u L_p},\mathbf{u}_f (t) \in \mathbb{R}^{n_u L_f}$ and the innovations $\mathbf{e}_p (t)\in \mathbb{R}^{n_y L_p},\mathbf{e}_f (t) \in \mathbb{R}^{n_y L_f}$ are column vectors defined similarly using these indices. 

Sequentially, we can write the state equation in (\ref{eq:pred_form}) as 
\begin{equation} \label{eq:est_x}
\begin{aligned}
    x(t) =& \sum_{k=0}^{L_p-1} \tilde{A}^k [ K y(t-k-1) + \tilde{B} u(t-k-1) ]  \\
    & + \tilde{A}^{L_p} x(t-p)
\end{aligned} 
\end{equation}
Assuming $\tilde{A}$ is stable, this implies that the state can be estimated by a linear combination of past inputs and outputs by choosing $L_p$ large enough such that $\tilde{A}^{L_p}$ can be ignored. Thus, we can replace the state by the estimate $\hat{x}(t)= \mathcal{K} \mathbf{z}_p(t)$, where $\mathcal{K}$ is a matrix of unknown coefficients and $\mathbf{z}_p(t)=[\mathbf{y}^\top_p(t), \mathbf{u}^\top_p(t)]^\top \in \mathbb{R}^{(n_u+n_y)L_p}$ is the column vector built from past data. 

The subspace predictor (SPC) in \cite{SPC} uses the innovation dynamics (\ref{eq:inno_form}) to establish the linear least square relation $\mathbf{y}_f \approx L [\mathbf{z}_p^\top , \mathbf{u}_f^\top ]^\top$. We instead use the predictor dynamics (\ref{eq:pred_form}) and arrive at the following alternative form
\begin{equation} \label{eq:column}
    \begin{aligned}
        \mathbf{y}_f(t) &= \Gamma x(t) + \Phi_u \mathbf{u}_f(t) + \Phi_y \mathbf{y}_f(t) + \mathbf{e}_f(t) \\
        & \approx \Gamma \mathcal{K} \mathbf{z}_p(t) + \Phi_u \mathbf{u}_f(t) + \Phi_y \mathbf{y}_f(t) + \mathbf{e}_f(t) ,
    \end{aligned}
\end{equation}
where
\begin{align}
    \Gamma &= \begin{bmatrix}
        C \\ C\tilde{A} \\ \vdots \\ C\tilde{A}^{L_f -1}
    \end{bmatrix} \\
    \Phi_u &= \begin{bmatrix}
        D & 0 & \cdots & & 0 \\
        C\tilde{B} & D & & &  \\
        C\tilde{A}\tilde{B} & C\tilde{B} & \ddots & & \vdots \\
        \vdots & \vdots & \ddots & &  \\
        C\tilde{A}^{L_f -2} \tilde{B} & C\tilde{A}^{L_f -3} \tilde{B} & \cdots & C\tilde{B} & D
    \end{bmatrix} \label{eq:phiu} \\
    \Phi_y &= \begin{bmatrix}
        0 & 0 & \cdots & &  0 \\
        CK & 0 & & & \\
        C\tilde{A}K & CK & \ddots & & \vdots \\
        \vdots & \vdots & \ddots & &  \\
        C\tilde{A}^{L_f -2} K & C\tilde{A}^{L_f -3} K & \cdots & CK & 0
    \end{bmatrix} \label{eq:phiy}
\end{align}
Observe that the coefficients in $\Phi_u$ and $\Phi_y$ are nothing but the predictor/observer Markov parameters, or the one-step-ahead predictor coefficients. 

We further define the Hankel matrices: 
\begin{align}
    Y_p & = \left[ \mathbf{y}_p(t_0),  \dots, \mathbf{y}_p(t_0+N-1) \right] \in \mathbb{R}^{n_yL_p \times N} \\
    Y_f & = [\mathbf{y}_f(t_0), \dots, \mathbf{y}_f(t_0+N-1)] \in \mathbb{R}^{n_yL_f \times N}  
\end{align}
the Hankel matrices $U_p \in \mathbb{R}^{n_uL_p \times N}$, $U_f \in \mathbb{R}^{n_uL_f \times N}$, $E_f \in \mathbb{R}^{n_yL_f \times N}$ and $Z_p = [Y_p^\top,U_p^\top]^\top \in \mathbb{R}^{(n_y+n_u)L_p \times N}$ are defined similarly. Then we can write (\ref{eq:column}) in the Hankel matrix form
\begin{equation}
    Y_f \approx \Gamma \mathcal{K} Z_p + \Phi_u U_f +\Phi_y Y_f +E_f
\end{equation}

\begin{remark}
    When using (\ref{eq:column}) as a predictor by setting $\mathbf{e}_f=\mathbf{0}$, we obtain a linear relationship between recent past input-output data and future predictions. This model resembles the linear relationship in Willems’ fundamental lemma, but it does not use raw data matrices as the model.
\end{remark}

\subsubsection{Receding-horizon control strategy}
As in standard MPC, DeePC, and DDPC formulations, all controllers here operate in a receding-horizon manner: at each time step $t$, a future input sequence $\mathbf{u}_f$ is optimized using a multi-step predictor $\hat{\mathbf{y}}_f(\mathbf{u}_f)$ by minimizing a given cost, such as a quadratic cost 
\begin{equation} 
\min_{\mathbf{u}_f} \quad \Vert \hat{\mathbf{y}}_f(\mathbf{u}_f)-\mathbf{r} \Vert^2_Q + \Vert \mathbf{u}_f \Vert^2_R
\end{equation}
where $\mathbf{r}$ is a reference signal. The optimization is done subject to the system equations and input/output constraints. Only the first control action is applied, and this procedure is repeated at every time step $t$. All methods in Section \ref{sec:experiment} share this same structure and differ only in the predictor used to generate $\hat{\mathbf{y}}_f$.

\section{the Subspace ARX (SSARX) Method} \label{sec:ssarx}

In this section, we propose our method for estimating the coefficients in (\ref{eq:column}). The estimations are divided into two stages. 

\textit{Stage 1: High-order ARX for $\hat{\Phi}_u$ and $\hat{\Phi}_y$}

We first estimate a high-order ARX to get an unstructured estimation of the impulse response coefficients. By choosing the ARX order to be $n_a$ in $y$ and $n_b$ in $u$, we get the ARX estimate 
\begin{equation} \label{eq:stage1-arx}
y(t) \approx \sum_{i=0}^{n_a-1} \hat{\phi}_{y,i} y(t-i) + \sum_{j=0}^{n_b-1} \hat{\phi}_{u,j}u(t-j) + e(t),
\end{equation}
where $n_a,n_b \geq L_f$ should be larger than the future Hankel dimension. The nominal observer/predictor Markov parameters can be written as 
\begin{equation} \label{eq:markov}
\begin{aligned}
    & \phi_{y,0}=0, \; \phi_{y,i}=C\tilde{A}^{i-1} K, \\
    & \phi_{u,0}=D, \; \phi_{u,j}=C\tilde{A}^{j-1} \tilde{B}
\end{aligned}
\end{equation}
for $i=1,\dots, n_a-1$ and $j=1,\dots, n_b-1$. 

From the estimated $\hat{\phi}_u$ and $\hat{\phi}_y$ coefficients in ARX, we form $\hat{\Phi}_u$ and $\hat{\Phi}_y$ in (\ref{eq:phiu}) and (\ref{eq:phiy}) by stacking them in the matrix sub-diagonals. 

We define the part of the future data that is only determined by the past to be $\overline{Y}_f \coloneq Y_f -\Phi_u U_f -\Phi_y Y_f$, then
\begin{equation} \label{eq:ybar}
    \hat{\overline{Y}}_f = Y_f -\hat{\Phi}_u U_f -\hat{\Phi}_y Y_f \approx \Gamma \mathcal{K} Z_p   +E_f
\end{equation}

\textit{Stage 2: Regression/ Low-rank Regression for $\widehat{\Gamma \mathcal{K}}$}

We view (\ref{eq:ybar}) as linear regression of $\hat{\overline{\mathbf{y}}}_f (t)$ onto the past data $\mathbf{z}_p(t)$. Since the innovation is uncorrelated with the past, we can compute the normal least squares regression
\begin{equation} \label{eq:LS}
\widehat{\Gamma \mathcal{K}} = \hat{\overline{Y}}_f Z_p^\top (Z_p Z_p^\top)^{-1}.
\end{equation}

An alternative is to add a low-rank constraint. We can select the rank $r$ and perform a reduced-rank regression
\begin{equation}
    \widehat{\Gamma \mathcal{K}} = \arg \min_{rank(\Gamma \mathcal{K})=r} \Vert \hat{\overline{Y}}_f - \Gamma \mathcal{K} Z_p \Vert^2_F .
\end{equation}
Let $S_{yy} = \frac{1}{N} Y_f Y_f^\top$, $S_{zz} = \frac{1}{N} Z_p Z_p^\top$ and $S_{yz} = \frac{1}{N} Y_f Z_p^\top$. We consider the singular value decomposition (SVD) of the whitened cross-covariance matrix
\begin{equation}
    S_{yy}^{-1/2} S_{yz} S_{zz}^{-1/2}=U \Sigma V^\top
\end{equation}
where $U$ and $V$ are orthogonal matrices, and $\Sigma=\text{diag}(\sigma_1, \dots, \sigma_q)$ is a diagonal matrix containing the singular values in non-increasing order. To obtain a rank-$r$ approximation, we retain only the first $r$ singular values and their corresponding singular vectors $U_r$ and $V_r$. $\Sigma_r$ is the $r\times r$ leading principal submatrix of $\Sigma$. The resulting low-rank estimate of $\Gamma \mathcal{K}$ is given by
\begin{equation} \label{eq:LS_LR}
    \widehat{\Gamma \mathcal{K}} = S_{yy}^{1/2} U_r \Sigma_r V_r^\top S_{zz}^{-1/2} .
\end{equation}
From this low-rank regression, we can incorporate structural prior (order constraints) without explicitly constructing a state-space realization. In practice, we can choose $r$ by checking the singular values.

The estimated multi-step ahead SSARX predictor of $\mathbf{y}_f(t)$ given past data and $\mathbf{u}_f(t)$ is thus the solution $\hat{\mathbf{y}}_f(t)$ of
\begin{equation} \label{eq:prediction}
    \hat{\mathbf{y}}_f(t) = \widehat{\Gamma \mathcal{K}} \mathbf{z}_p(t) + \hat{\Phi}_u \mathbf{u}_f(t) + \hat{\Phi}_y \hat{\mathbf{y}}_f(t) .
\end{equation}
Algorithm \ref{alg:ssarx-spc} summarizes the SSARX identification step used to construct the multi-step predictor.

\begin{algorithm}[t] \label{algo}
  \caption{SSARX identification steps}
  \label{alg:ssarx-spc}
  \begin{algorithmic}[1]
    \State Collect open- or closed-loop data $\{u_t,y_t\}_{t=1}^N$.
    \State Choose $L_p,L_f$, the ARX orders $n_a,n_b$.
    \State Build $U_p,U_f,Y_p,Y_f$ and set $Z_p = [Y_p^\top, U_p^\top]^\top$.
    \State Estimate a high-order ARX model to get the predictor Markov parameters $\hat{\phi}_u, \hat{\phi}_y$, and stack in matrix sub-diagonals to get the estimated $\hat{\Phi}_u, \hat{\Phi}_y$.
    \State Compute $\hat{\overline{Y}}_f = Y_f - \hat{\Phi}_u U_f - \hat{\Phi}_y Y_f$. Then regress $\hat{\overline{Y}}_f$ on only the past data $Z_p$ to get $\widehat{\Gamma\mathcal K}$, either via least-squares (\ref{eq:LS}) or low-rank regression (\ref{eq:LS_LR}).
    \State Use (\ref{eq:prediction}) for multi-step ahead prediction. 
  \end{algorithmic}
\end{algorithm}

\subsubsection{Causality}
The predictor (\ref{eq:prediction}) is causal because of the lower-triangular structure of $\hat{\Phi}_u$ and $\hat{\Phi}_y$ matrices, i.e., the $i$-th block component only depends on $u$ and $y$ up to time $i-1$. This contrasts with the fundamental lemma, where the predictor essentially uses non-causal Hankel coordinates and needs explicit causality-enforcing tricks (\cite{Sader2025}). 

\subsubsection{Closed-loop consistency}

Under standard assumptions on the closed-loop experiment and the persistency of excitation of the input, the SSARX estimator in (\ref{eq:prediction}) yields a consistent stacked multi-step predictor.

For strictly proper system ($D=0$) and sufficiently large ARX orders $n_a,n_b$, the high-order ARX estimator in (\ref{eq:stage1-arx}) provides consistent estimates of the one-step-ahead predictor coefficients $\hat{\phi}_{y,i}, \hat{\phi}_{u,j}$, and thus of the corresponding predictor/observer Markov parameters collected in $\hat{\Phi}_y$ and $\hat{\Phi}_u$; see e.g., \cite{Ljung1992} and \cite{WNSF} for closed-loop ARX consistency results.

As a consequence, for a large enough $L_p$, the future block $\hat{\overline{Y}}_f$ in (\ref{eq:ybar}) converges to its ideal counterpart $\overline{Y}_f$ as the sample size $N$ increases. For a fixed finite prediction horizon $L_f$, the relation $\overline{Y}_f=\Gamma \mathcal{K} Z_p+E_f$ is a standard linear regression of the stacked future outputs onto the past data $Z_p$, where the innovation $E_f$ is uncorrelated with $Z_p$. Under the same persistence-of-excitation assumption on $Z_p$, the least square estimator in (\ref{eq:LS}) is therefore consistent, and $\widehat{\Gamma \mathcal{K}} \to \Gamma \mathcal{K}$ as the data length $N\to \infty$. 

Combining these two steps shows that the SSARX predictor in (\ref{eq:prediction}) converges to the true stacked multi-step predictor for any fixed $L_f$, i.e., the SSARX-based multi-step predictor is closed-loop consistent. This property is also mentioned in \cite{ChiusoConsistency}. It contributes to the bias removal compared with SPC, see e.g. \cite{bias}.



\section{Experiment} \label{sec:experiment}

In this section, numerical simulations are carried out to evaluate the control performance and the bias of the SPC-type predictive controllers. We consider a sampled second-order system presented in \cite{innovation} and \cite{Breschi2023}, with all experimental settings identical to those in \cite{innovation}. The LTI system can be described by 
\begin{equation} 
\begin{cases}
x(t+1) = A x(t) + B u(t) + w(t), \\
y(t) = C x(t) + D u(t) + v(t), 
\end{cases}
\end{equation}
with the system matrices
\begin{equation}
    \begin{aligned}
        A& = \begin{pmatrix}
            0.7326 & -0.0861     \\
            0.1722 & 0.9909
        \end{pmatrix}, \quad B  = \begin{pmatrix}
        0.0609 \\ 0.0064 \end{pmatrix} , \\
        C & = \begin{pmatrix}
            0 & 1.4142 \end{pmatrix}, \quad D=0 ,
    \end{aligned}
\end{equation}
the process noise variance $\Sigma_w = \sigma_w^2 \mathbb{I}_2$, and the measurement noise variance $\Sigma_v = \sigma_v^2$. The control objective is to make the output track a reference signal $r(t)$ while satisfying the input and output constraints $-2 \leq u \leq 2, -2 \leq y \leq 2$. 

For closed-loop training data collection, we generate a trajectory of length $N_{\text{train}} = 200$ using the simple feedback $u(t) = r_{\text{train}}(t)-y(t)$, where the external reference $r_{\text{train}}(t)$ is a square wave with a period of $50$ time-steps and amplitude of $2$, contaminated by a zero-mean Gaussian distributed sequence with variance $0.01$. The past and future horizons are $L_p=10$ and $L_f=15$. To evaluate the control performance, we use the cost function $J = \sum_{t=1}^{N_{\text{test}}} ||y(t)-r(t)||_Q^2 + ||u(t)||_R^2$, where $Q = 1$ and $R = 0.01$. Each test trajectory length is $N_{\text{test}} = 100$. To get a quantitative measure of the noise level, we define $\text{SNR}=10 \log_{10} \frac{P[\mathbf{y}^0(t)-\mathbf{y}(t)]}{P[\mathbf{y}^0(t)]}$, where $P[\cdot]$ denotes the signal power, and $\mathbf{y}^0(t)$ is the ideal output without any noise. It is the ratio of noisy output power to noise-free output power, expressed in dB.

We investigate the performance of the following predictors:

(1) \textbf{SPC} The original Subspace Predictive Controller based on regression in \cite{SPC}.

(2) \textbf{CLSPC} Closed-loop SPC based on \cite{CLSPC}.

(3) \textbf{IV-DDPC} Instrumental variable based on (20) in \cite{IVDDPC}, with left coprime factorization (LCF) of the controller and the reference as instruments.

(4) \textbf{InnoOP} DDPC with augmented innovation estimation in \cite{innovation}, using an ARX order $\rho=15$ as in the original paper.

(5) \textbf{SSARX} Our proposed SSARX predictor with LS regression in (\ref{eq:LS}) and (\ref{eq:prediction}), with the ARX order equal to the future Hankel dimension $n_a=n_b=L_f$. 

(6) \textbf{SSARX-LR} Our proposed SSARX predictor with low-rank regression (\ref{eq:LS_LR}) and (\ref{eq:prediction}), with the same ARX order $n_a=n_b=L_f$ and the reduced rank $r=2$. 

(7) \textbf{MPC-SSKF} Oracle MPC with true parameters, state estimation using steady state Kalman filter. Baseline performance.

\begin{table}[tb] \label{tb:1}
\begin{center}
\caption{Measurement noise $\sigma_v$ and
process noise $\sigma_w$ ($\times 10^{-2}$)} \label{tb:setting}
\begin{tabular}{cccc} \hline
SNR  & (1) $\sigma_v/\sigma_w $ & (2) $\sigma_v/\sigma_w $ & (3) $\sigma_v/\sigma_w $ \\ \hline
30dB & 1.3/0 & 0.75/0.187 & 0.2/0.25 \\
25dB & 2.3/0 & 1.5/0.37 & 0.2/0.50 \\ 
20dB & 4.2/0 & 2.5/0.62 & 0.2/0.89 \\ 
15dB & 7.4/0 & 4.5/1.13 & 0.2/1.37 \\ \hline
\end{tabular}
\end{center}
\end{table}

\begin{figure}[tb]
\begin{center}
\includegraphics[width=8.4cm]{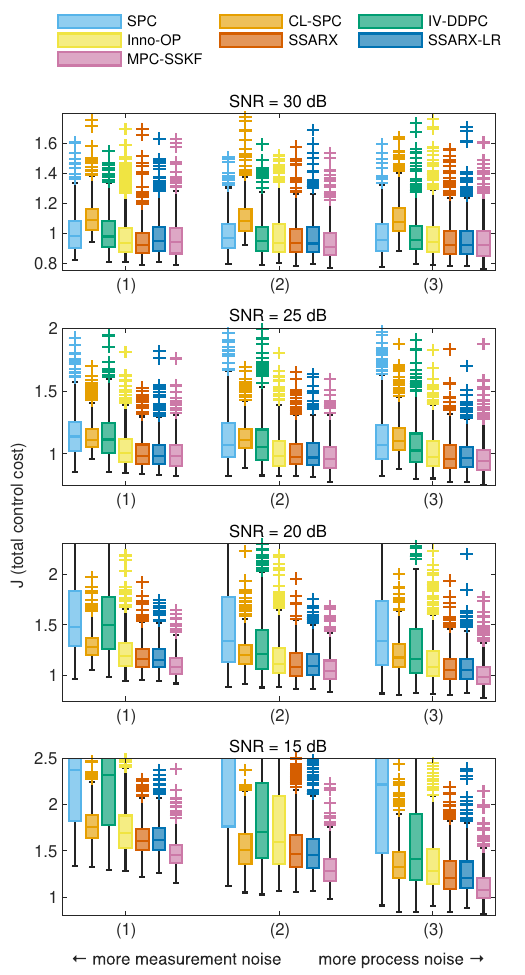}    
\caption{Difference in control cost for different SNR (30dB, 25dB, 20dB, 15dB from top to bottom). From Left to right, group (1) has only measurement noise, and group (3) has mainly process noise.} 
\label{fig:fig1}
\end{center}
\end{figure}

\subsection{Measurement and process noise}

We want the controlled output to follow a sinusoid reference signal $r(t)=\sin (2\pi t/N_{\text{test}})$. To show the influences of the process noise $\sigma_w$ and the measurement noise $\sigma_v$, we change their ratio while keeping the same SNR. For each SNR, we have (1) only measurement noise $\sigma_v$, (2) $\frac{\sigma_v}{\sigma_w} \approx 4$, (3) small measurement noise $\sigma_v$ and high process noise $\sigma_w$, as in Table 1.

We test the controller by generating different training and testing data. Fig.~\ref{fig:fig1} shows the difference in control cost over $N_{MC}=500$ Monte Carlo simulations.

From Fig.~\ref{fig:fig1}, we can observe that these closed-loop methods respond differently to measurement and process noise, even if the SNR level is the same. The proposed SSARX and low-rank regression SSARX perform well among these methods.

\begin{remark}
    All methods listed here have relatively low computational cost compared to the fundamental lemma-based controllers, in that they do not have to solve for the additional vector $g$ in each iteration. 
\end{remark}

\begin{figure}[tb]
\begin{center}
\includegraphics[width=8.4cm]{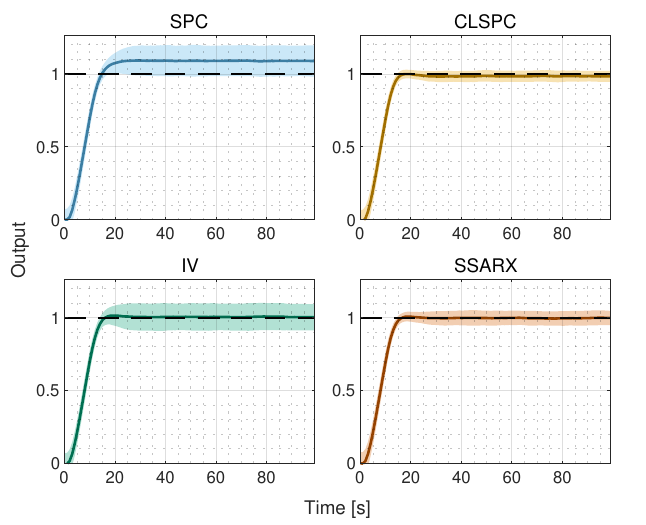}    
\caption{Controlled output trajectory when the reference is constant, with the mean and standard deviation (shaded areas) over 500 Monte Carlo runs. Training data length $N_{train}=200$, noise level SNR=20dB with high process noise. We use the stationary part $t \geq 50$ to compute the bias and variance in Fig.~\ref{fig:fig3}. } 
\label{fig:fig2}
\end{center}
\end{figure}

\begin{figure}[tb]
\begin{center}
\includegraphics[width=8.4cm]{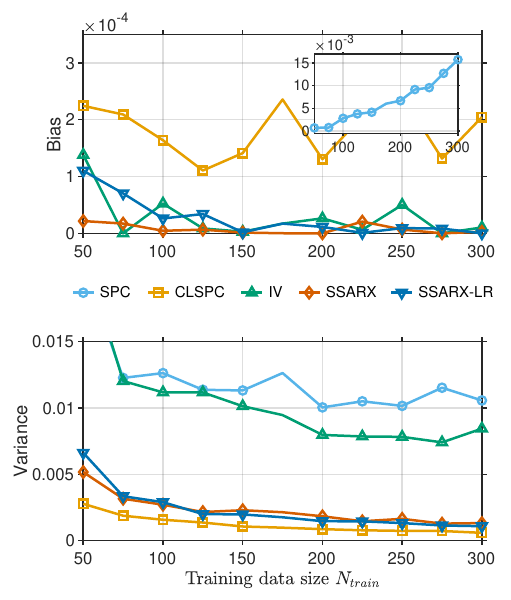}    
\caption{Bias (upper) and variance (lower) as a function of training data size $N_{train}$, when the controlled system is tracking a constant signal. Each point is computed from (\ref{eq:bias_var}) with 500 Monte Carlo simulations.} 
\label{fig:fig3}
\end{center}
\end{figure}

\subsection{Bias and variance}

We analyze the bias and variances for the SPC-type controllers. All these methods use an explicit multi-step predictor rather than the fundamental-lemma predictor, so we refer to them as SPC-type. We pick the scenario when SNR=20dB, $\sigma_v=0.2 \times 10^{-2}$ and $\sigma_w=0.89 \times 10^{-2}$. We set the reference signal to be the constant $r(t)=1$ for the whole test trajectory $N_{test}=100$. 

Fig.~\ref{fig:fig2} shows the controlled output trajectories from different controllers, with their mean and standard deviation over 500 Monte Carlo runs. SPC has a clear non-zero bias, which is also analyzed in \cite{bias}. 

To further check the bias and variance, we pick the stationary test trajectory part $50\leq t\leq N_{\text{test}}=100$. For the $n$-th Monte Carlo run, we define the time-averaged tracking error to be $e_{n}=\frac{1}{50} \sum_{t} [ y(t)-r(t) ]$. Let $\bar{e} = \frac{1}{N_{MC}}\sum_{n=1}^{N_{MC}} e_{n}$ be the mean error across Monte Carlo runs. Then we can define the scalar bias and variance to be
\begin{equation} \label{eq:bias_var}
    \text{Bias} = ||\bar{e}||_2^2 , \quad \text{Var} = \frac{1}{N_{MC}-1} \sum_{n=1}^{N_{MC}} || e_n - \bar{e} ||_2^2
\end{equation}

For this noise level, we change the length of the training data $N_{\text{train}}$ and observe the change in bias and variance. Fig.~\ref{fig:fig3} shows the comparison of different SPC-type predictive controllers. 

From Fig.~\ref{fig:fig3}, the SPC controller has a non-trivial bias for closed-loop training data, and the other closed-loop compatible methods reduce this bias. As $N_{\text{train}}$ increases, the bias of IV-DDPC, SSARX, and low-rank SSARX decreases. For SSARX, this systematic decay aligns with the closed-loop consistency result in Section \ref{sec:ssarx}, which states that, for sufficiently large ARX orders, the estimated multi-step predictor converges to the true ones as more closed-loop data are used.

For the variance analysis, the ARX-based methods CLSPC, SSARX, and low-rank SSARX have the lowest tracking variance.

\section{Conclusion}
SSARX provides a non-parametric, causal, closed-loop consistent predictor that fits directly into DDPC/MPC without fundamental lemma constraints.

Empirically, SSARX and its low-rank variant are among the best performers in terms of bias and variance on closed-loop data with significant process noise.

However, SSARX still requires persistently exciting data. Future work may incorporate prior information to mitigate the effects of limited excitation.

\bibliography{ifacconf}             

\appendix
\end{document}